\documentclass{aa}  

\usepackage{graphicx}
\usepackage{txfonts}
\def\kms {km\,s$^{-1}$ }
\begin{document}

   \title{The variability  of Betelgeuse explained by surface convection\thanks{Based on observations obtained at the T\'elescope \textit{Bernard Lyot}
   (TBL) at Observatoire du Pic du Midi, CNRS/INSU and Universit\'e de
   Toulouse, France.}}

    \author{{ Q.~Pilate}\inst{1},{ A.~López Ariste}\inst{1},{ A.~Lavail}\inst{1},{ Ph. Mathias}\inst{2} }

   \institute{
            Institut de Recherche en Astrophysique et Plan\'etologie,
            Universit\'e de Toulouse, UPS, CNRS, IRAP/UMR 5277,
            14 avenue Edouard Belin, F-31400, Toulouse, France  
            \and  IRAP, Universit\'e de Toulouse, CNRS, UPS, CNES, 57 avenue d'Azereix, 65000, Tarbes, France
             }

   \date{Received ...; accepted ...}

% \abstract{}{}{}{}{} 
% 5 {} token are mandatory
 
  \abstract
  {Betelgeuse is a red supergiant (RSG) that is known to vary semi-regularly on both short and long timescales. The origin of the short period of Betelgeuse has often been associated to radial pulsations but could also be due to the convection motions present at the surface of RSGs.}
   { We investigate the link between surface activity and the variability of the star.   }
   {Linear polarization in Betelgeuse is a proxy of convection which is unrelated to pulsations. Using  10 years of spectropolarimetric data of Betelgeuse, we seek for periodicities in the least-squares deconvolution profiles of Stokes $I$, $Q$, $U$ and the total linear polarization using  Lomb-Scargle periodograms. }
   {We find similar periods in linear polarization signals than in photometric variability. The 400 d period is too close to a peak of the window function of our data. But the two periods of 330 d and 200 d are present in the periodogram of Stokes $Q$ and $U$, showing that  the variability of Betelgeuse can be interpreted as due to  surface convection. }
   {Since linear polarization in the spectrum of Betelgeuse is not known to vary with pulsations, but is linked to surface convection, and since similar periods are found in time series of photometric measurements and spectropolarimetry, we conclude that the photometric variability is due to the surface convective structures, and not to any pulsation phenomenon.}


   \maketitle
%
%-------------------------------------------------------------------

\section{Introduction}
\label{sec: Introduction}
Betelgeuse is a prototypical red supergiant (RSG), known to be a semi-regular variable. Several periods can be found in the 
literature, usually clustered around the so-called long secondary period (LSP) of 2000 d, plus shorter periods around 400 and 200 d \citep{kiss_variability_2006}. The origin of the LSP remains a source of discussion: some have tried to find the explanation of the LSP in the lifetime of the large convective cells present at the surface of RSGs \citep[e.g][]{stothers_giant_2010} while others have invoked a magnetic field \citep{wood_long_2004}. The shorter periods of 400 and 200 d are given a different origin than the LSP, and are traditionally attributed to radial pulsation, the 400 d period  being the fundamental mode and the 200 d the first overtone \citep{Guo_2001,joyce_standing_2020}. By studying a large sample of RSGs, \cite{kiss_variability_2006} found a period of 388 $\pm$ 30 d, which was later confirmed by \cite{chatys_periodluminosity_2019}. Further works have been carried out to identify periods in Betelgeuse, for instance, using the observations from the AAVSO database between 1995 and 2014, \cite{montarges_close_2016} found a period of 423.59 $\pm$ 39 d. Furthermore, \cite{jadlovsky_analysis_2023} found a photometric period of 417 $\pm$ 17 d, while \cite{joyce_standing_2020} found a period of 416 $\pm$ 24 d and identified another shorter period of 185 $\pm$ 13.5 d being attributed to the first overtone, in agreement with the historical work of \cite{stothers_pulsation_1969}.\\

Besides radial pulsation, surface activity leading to random brightness variation has also been mentioned to explain the variability of Betelgeuse, first theorized by \cite{schwarzschild_scale_1975}. More recently, \cite{gray_mass_2008} proposed that the 400 d period was a consequence of convective activity. Betelgeuse is known to present large convective cells with lifetimes of the order of one to two years \citep{lopez_ariste_convective_2018} and measurable changes 
within the span of one week. Those typical convective timescales are also seen in numerical simulations of RSGs, where the smallest convective granules live on a timescale of a few weeks/month and the largest on a timescale of a few months up to a few years \citep{chiavassa_probing_2022}. Bright convection cells near the disk center will increase the integrated brightness of the star compared with other situations where such cells are found near the edges. So, even without invoking the formation of 
dust or other changes in opacity, one may argue that the simple evolution of convective patterns of the star may create variability. Such 
variability could be expected to be random, but will present quasi-periodicities related to the typical time scales of the 
convective patterns \citep{gray_mass_2008}. \\

At the end of 2019, Betelgeuse reached a historical minimum in its luminosity, called the Great Dimming \citep{guinan_fall_2020}. 
From interferometric and spectroscopic data, it has been proposed that this event was caused by the formation of a cloud of dust close to the line of sight \citep{montarges_dusty_2021}. 
Interferometric images show a drop in luminosity in the southern hemisphere of Betelgeuse, which could be caused by a mass loss event and leading to this dimming \citep{dupree_great_2022}. Other hypotheses have also been explored, such as a drop in temperature \citep{harper_photospheric_2020}, or an increase in the molecular opacity \citep{kravchenko_atmosphere_2021}. In this event, it turns out that a change in brightness was not due to pulsation but to a change in the 
brightness distribution over the stellar disk.
Since the end of the Great Dimming event, Betelgeuse has continued its random variation of brightness. 
But quite interestingly, 
it has been shown by \cite{jadlovsky_analysis_2023} and \cite{dupree_great_2022} that the periodicity of Betelgeuse has changed since the dimming. Using the light curves from AAVSO, 
the cited authors showed that before the dimming, the dominant period of Betelgeuse was the 400 d period, while after the great dimming, the main periods of Betelgeuse have shortened, oscillating between 97 d and 230 d \citep{dupree_great_2022}, revealing a change in the behavior of the atmosphere of Betelgeuse.
While the great dimming may be seen as 
a singular event, such modifications of the variability periods bring up the question of whether all changes in brightness can be due to similar changes in the brightness 
distribution over the disk and unrelated to any pulsation phenomenon. 
Our interest in an alternative explanation of the variability
in terms  of convective patterns arose. In order to address this question, we seek such typical periods of 400 and 200 d in  
observational proxies related to the convective
activity but not to any pulsation, such as the linear polarization
spectra.\\

Linear polarization in the atomic lines of the spectrum of Betelgeuse, discovered by \cite{auriere_discovery_2016}, has been interpreted as the joint action of 
two mechanisms. First, the depolarization of the continuum by atoms which absorb linearly polarized light from the continuum and re-emit unpolarized light,
the continuum photons being polarized by Rayleigh scattering. This depolarization produces signals with azimuthal symmetry over the visible disk, which would  cancel out the net linear polarization on a homogeneously bright disk. Thus, such mechanism of polarization must be combined with an inhomogeneously bright disk to produce the net linear polarization 
signal observed. This interpretation suggested the possibility of mapping those brightness inhomogeneities. This has been achieved by \cite{lopez_ariste_convective_2018}, who produced even 3-dimensional images of the atmosphere of Betelgeuse \citep{lopez_ariste_three-dimensional_2022} by taking advantage of the different heights of formation of 
different lines in the spectrum of Betelgeuse. The produced images compare very well with contemporaneous images made with interferometric 
techniques \citep{montarges_close_2016} and show clear convective patterns, akin to solar granulation.\

Linear polarization observed in the atomic lines is therefore a proxy of convection, a priori unrelated to radial 
pulsations but linked to the brightness inhomogeneities due to the convective patterns in Betelgeuse. If one can find the aforementioned variability periods 
in linear polarization, we can conclude that these periods are related to the convective activity which originates the linear polarization signals.
This is the purpose of the present work.\

In section \ref{Section 2}, we describe the dataset of linear polarization obtained with Narval and Neo-Narval at the 2-meter Télescope \textit{Bernard Lyot} (TBL), as well as the Lomb-Scargle periodograms used to derive periods. In section \ref{section 3}, we seek periods in the Least-Squares Deconvolution (LSD) profiles of Betelgeuse using the Lomb-Scargle periodogram. In section \ref{section 4}, we associate the 200 d period with the convective timescale of the smallest granules and the 330 d periodicity with the largest granules. We speculate on an explanation in the change of variability of Betelgeuse before and after the great dimming.

\section{The polarimetric data}
\label{Section 2}

Betelgeuse has been observed since 2013 with Narval and Neo-Narval at the Télescope \textit{Bernard Lyot}\footnote[1]{https://tbl.omp.eu/}. The polarimeter Narval observed Betelgeuse on average every month from November 2013 to April 2019, except during summer, representing a total of 56 observations over 5.5 years. During this period, Betelgeuse was observed 43\% of the time between January and April and 57\% between September and December. Before the great dimming of Betelgeuse, Narval was replaced by Neo-Narval which is now observing Betelgeuse since early 2020. These two instruments measure the polarization 
over the visible and near infrared spectra of Betelgeuese (390-1000 nm) with high spectral resolution (R=65000) and high polarimetric sensitivity. The signal-to-noise ratios are not sufficient to measure the weak polarization signals in individual atomic lines of the 
spectrum. These amplitudes are known to be of the order of $10^{-4}$ times the continuum intensity, and they only exceptionally reach amplitudes of $10^{-3}$. When these high amplitudes are present, enough photons can be accumulated per spectral bin to 
detect the linear polarization signal above noise in lines \citep{auriere_discovery_2016}. Most commonly, the amplitudes of linear polarization are below noise levels. On those 
occasions, we have to add up the signals of thousands of lines to increase the signal-to-noise ratios. This addition 
is performed through the Least-Squares Deconvolution \citep[LSD;][]{donati_spectropolarimetric_1997,kochukhov_least-squares_2010}.
This technique has been successfully used in the past to measure 
magnetic field distributions over stellar surfaces \citep[e.g][]{donatiLandstreet2009,reiners2012,kochukhov2021} and is now employed to produce images of the brightness distributions in the photosphere 
of Betelgeuse and other RSGs, as cited in section~\ref{sec: Introduction}. To produce the LSD profiles, we used a solar abundance line mask, similar to the one used by \cite{auriere_discovery_2016} and \cite{mathias_evolution_2018}, which was calculated from the data provided by VALD \citep{kupka_vald2_1999}, for an effective temperature of 3750 K, $\log g =0.0$ and a microturbulence of 4 \kms, aligning with the physical parameters of \cite{josselin_atmospheric_2007}. The mask includes approximately 15\,000 lines with depths exceeding 40\% of the continuum.

In this work, we shall examine the signals produced through the LSD technique, which produce a pseudo-spectral line in intensity 
and linear polarization. This pseudo-spectral line does not correspond to any particular atomic species but rather represents an average of all species present and emitting in the 
photosphere of Betelgeuse. Following these procedures of line addition, the resulting profiles carry the coherent signals present in the photospheric lines, but the specific characteristics of individual spectral lines are erased. All of this data has been previously presented before by \cite{auriere_discovery_2016}, \cite{mathias_evolution_2018} and 
\cite{lopez_ariste_three-dimensional_2022}, which also provide details on observing times and conditions as well as a more detailed description 
of the data reduction process\footnote[2]{Beyond a 2-year proprietary embargo, and up to technical issues, all these data is available through 
PolarBase (http://polarbase.irap.omp.eu/).}.\\
Note that all the velocities mentioned in the text or in the figures are provided in the heliocentric frame.

\section{Period search}
\label{section 3}

Attempting to identify periods in an astrophysical context faces the challenge of unevenly spaced observed data. To overcome this hurdle, 
we used the Lomb-Scargle periodogram \citep{lomb_least-squares_1976,scargle_studies_1982} as implemented in astropy \citep{the_astropy_collaboration_astropy_2018}. This method involves fitting sinusoids to the data using least-squares with frequency sampled between the first and last observation. The quality of the fit determines the power attributed to a frequency. 
We applied this technique to the LSD profiles of Stokes $I$, $Q$, $U$ and to the total linear polarization of Betelgeuse observed by the TBL with Narval from 2013 to 2019, with a total of 56 observations. We also looked at the periodograms obtained from the dataset available post-dimming, from 2020 to 2024. However, the base time of this dataset is not long enough to produce meaningful periodograms. Therefore, our focus was on periodograms computed exclusively from data obtained before the great dimming with Narval.

\subsection{Variability of the Stokes parameters}

We first computed the Lomb-Scargle periodogram at each wavelength of the LSD profile, ranging from -25~\kms to +60~\kms with a step of 1.8~\kms to cover the entire 
signal present in the LSD profile. \cite{lopez_ariste_convective_2018} identified the most blueshifted signal towards -20~\kms and interpreted it as  the maximum velocity of the rising plasma. The most redshifted signal was often found at +40~\kms, interpreted 
as the rest velocity of the star. Sometimes, signals at velocities greater than +40~\kms can be found, with peak in the linear polarization signal up to 50~\kms \citep{lopez_ariste_three-dimensional_2022}, which can be interpreted either as dark and cold plasma falling back to Betelgeuse or as rising plumes of plasma located on the hidden face of Betelgeuse, ascending so high that they appear above the limb \citep[for the case of the RSG $\mu$~Cep]{lopez_ariste_height_2023}. The +60~\kms limit in the Lomb-Scargle periodograms was set to include the occasional linear polarization peaks beyond the velocity of the star.

\begin{figure}[!h]
    \centering
    \includegraphics[width=0.5\textwidth]{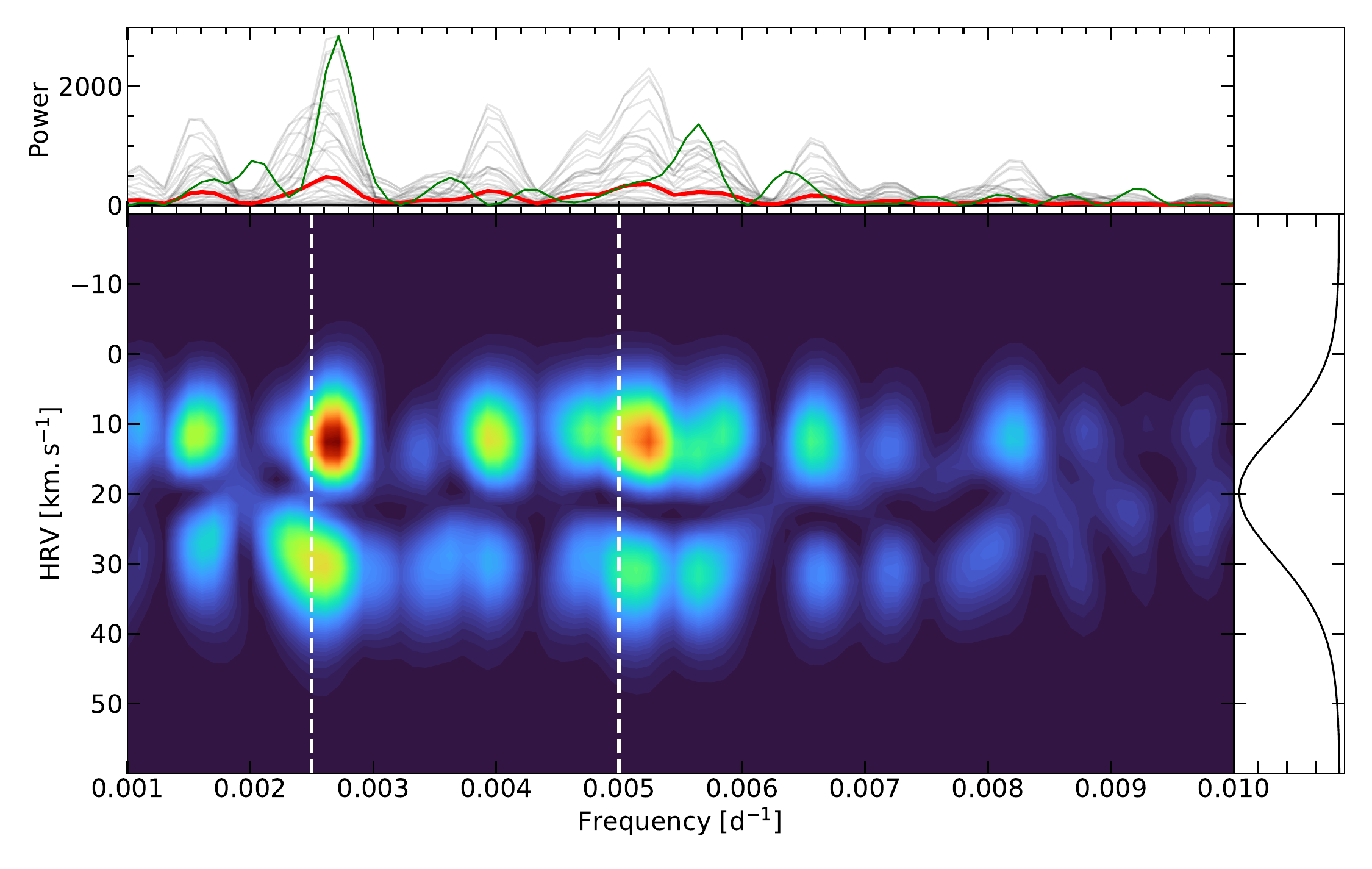}
    \caption{Lomb-Scargle periodogram of the LSD profile of intensity.
    The upper panel is the Lomb-Scargle periodogram for each velocity bin (gray lines) and the average (red line). The green line is the window function.
    In the lower panel, the two white dashed lines mark respectively the 400 d and the 200 d periods. The y-axis represents the heliocentric radial velocity (HRV). The color axis marks the power of the fit, red shows a higher power than dark blue. The right panel is the average intensity profile for each velocity bin.}
    \label{LS intensity}
\end{figure}

\begin{figure}[!h]
    \centering
    \includegraphics[width=0.5\textwidth]{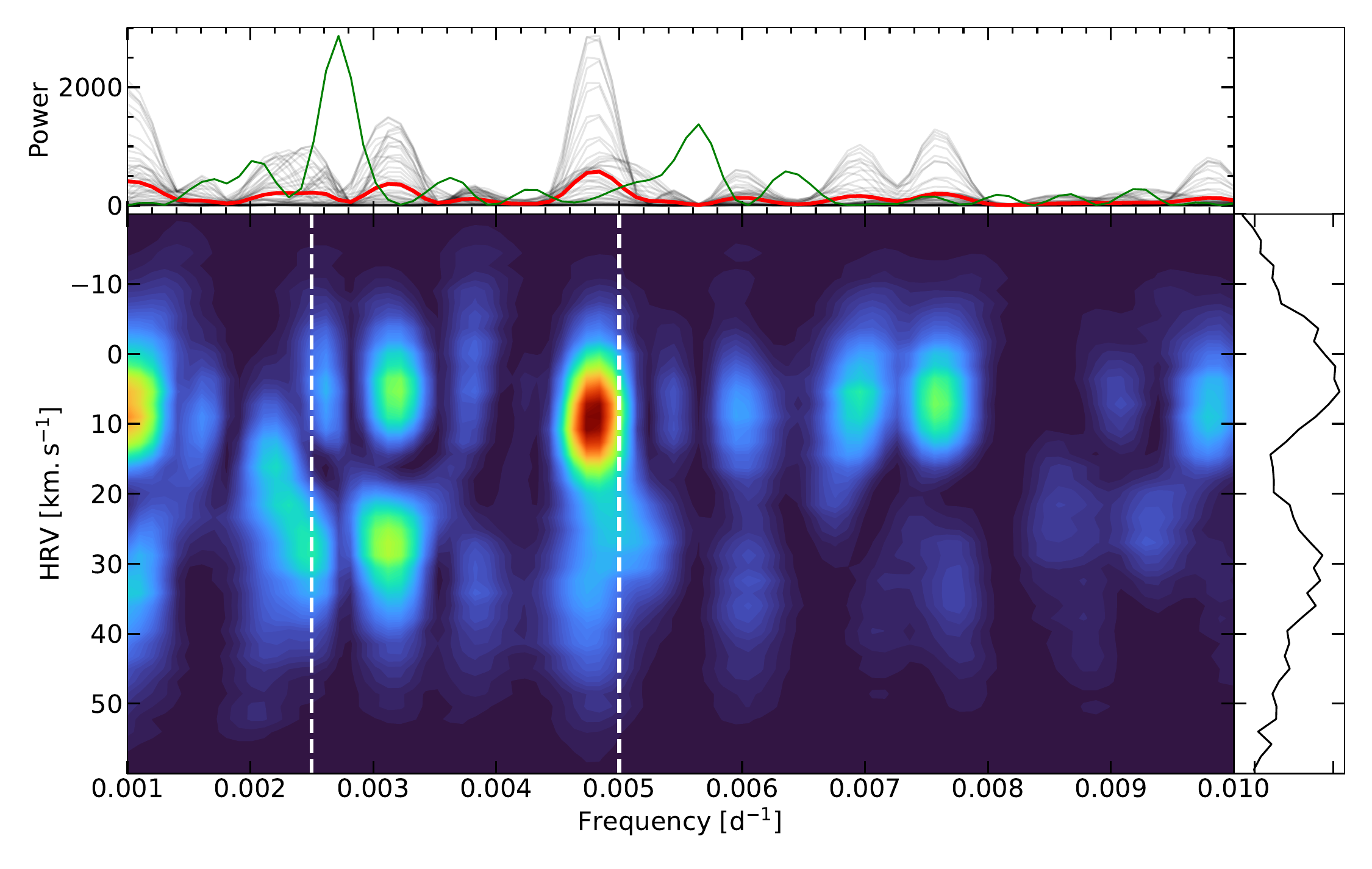}
    \caption{Same as Fig. \ref{LS intensity} for Stokes $Q$. }
    \label{LS Q}
\end{figure}

\begin{figure}[!h]
    \centering
    \includegraphics[width=0.5\textwidth]{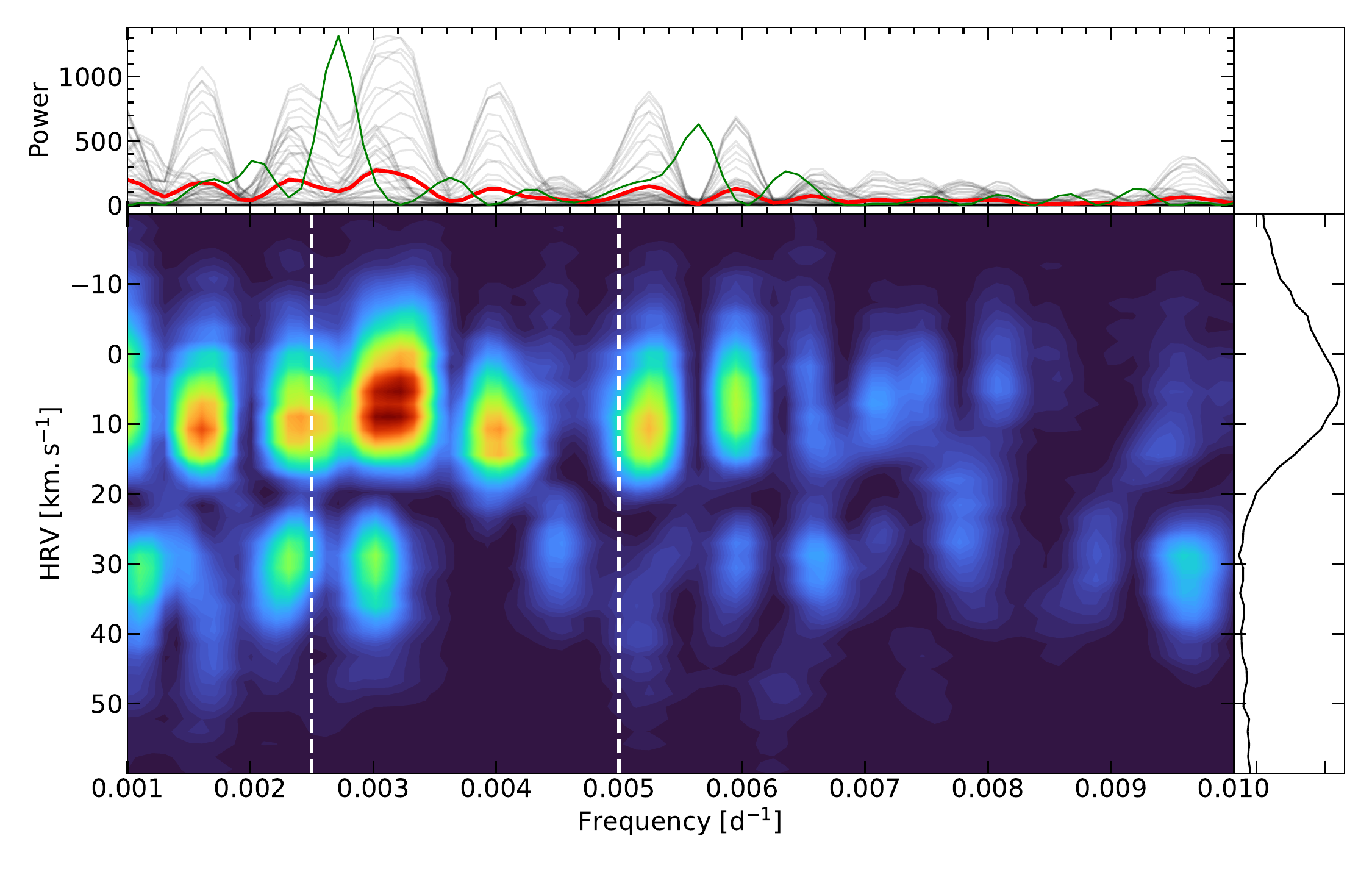}
    \caption{Same as Fig. \ref{LS intensity} for Stokes $U$.}
    \label{LS U}
\end{figure}

\begin{figure}[!h]
    \centering
    \includegraphics[width=0.5\textwidth]{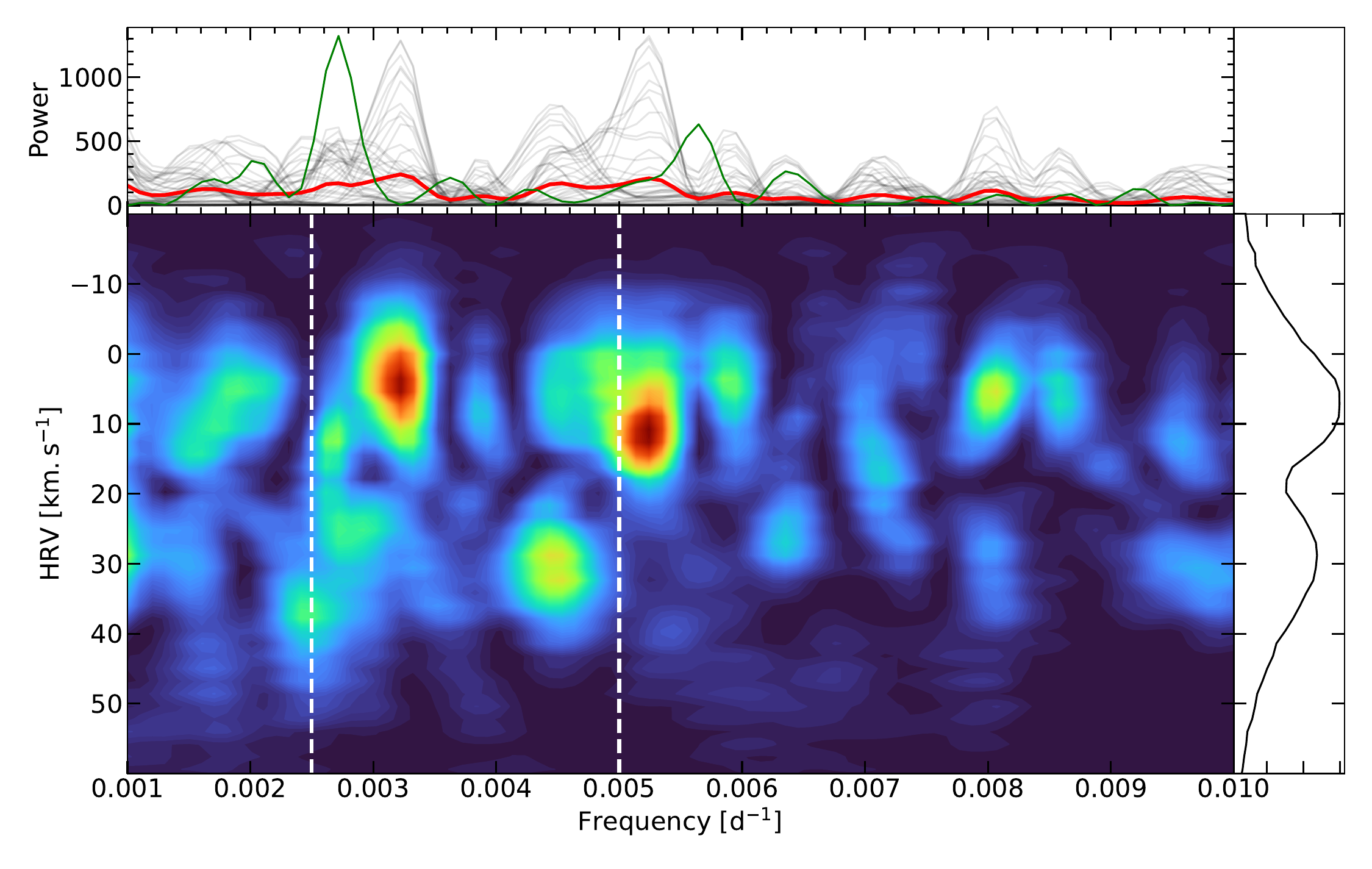}
    \caption{Same as Fig. \ref{LS intensity} for the total linear polarization: $\sqrt{Q^2+U^2}$.}
    \label{LS linear polarization}
\end{figure}

Figures \ref{LS intensity},\ref{LS Q},\ref{LS U} and \ref{LS linear polarization} show the Lomb-Scargle periodogram of the LSD profile for Stokes $I$, Stokes $Q$, Stokes $U$ and the total linear polarization, respectively. In each figure, the upper panel depicts the Lomb-Scargle periodogram of each velocity bin (gray lines) along with the total average (red line). The green line represents the window function. The spectral window reﬂects the pattern caused by the structure of gaps in the time sampling. The peaks pointed out by this window cannot be interpreted as ones related to the star. The lower panel shows the Lomb-Scargle periodogram for each velocity bin, with the white dotted lines marking the 400 and 200 d periods for reference. The right panel represents the mean profile of each Stokes parameter at each velocity bin.

Examining Fig.\ref{LS intensity}, which shows the Lomb-Scargle periodogram of the LSD of the intensity profile, we observe that the 400 and 200 d periods seem to be captured by the periodogram. However, the 400 d period is uncomfortably close to the peak of the window function. Furthermore, both periods are located at the same HRV and in the blue wing of the profile. Interestingly, \cite{mathias_evolution_2018} previously found this 200 d periodicity in spectroscopic observation, despite of a shorter observation period.

Figures \ref{LS Q} and \ref{LS U} display the Lomb-Scargle periodograms of Stokes $Q$ and $U$. These periodograms exhibit significant differences compared to the intensity one. In Fig. \ref{LS Q}, a prominent signal is evident around 200 d. Regarding Stokes $U$ in Fig. \ref{LS U}, it appears that the primary frequency is approximately $0.003 \ \mathrm{d^{-1}}$ (equivalent to a 330 d period). Other periods, such as those at 200 d or 250 d ($0.004 \ \mathrm{d^{-1}}$) are present, but are difficult to trust. From both periodograms of Stokes $Q$ and $U$, we recover the 200 d period, and also the 330 d period is notable, which aligns closely with the 400 d period reported in the literature and is in agreement with the timescale of the hysteresis loop reported by \cite{kravchenko_tomography_2019}.  
Figure \ref{LS linear polarization} shows the Lomb-Scargle periodogram of $\sqrt{Q^2+U^2}$, representing the total linear polarization of Betelgeuse. 
This periodogram confirms the significant powers at both 330 d and 200 d, consistent with the periodograms of Stokes $Q$ and $U$.

\subsection{Variability of the polarimetric imaging}

Using linear polarization, \cite{lopez_ariste_convective_2018} successfully reconstructed images of Betelgeuse, which have been compared favorably to inteferometric images obtained by \cite{montarges_close_2016}. The images are produced by finding the brightness distribution that better fits the observed linear 
polarization LSD profile using a Marquardt-Levenberg minimisation.

Betelgeuse is observed, on average, every month by the TBL, enabling the tracking of its surface activity through this technique of polarimetric imaging. This technique has previously allowed for the estimation of  the size and the
 lifetime of convective cells on the surface \citep{lopez_ariste_convective_2018}. From these images, we computed a photo-center, a quantity sensitive to the size and number of convective cells. The photo-center was computed from the equations 2 and 3 given by \cite{chiavassa_probing_2022}.
A homogeneous star will have a photo-center displacement coinciding with the barycenter of the star, whereas a star with one or two large convective cells will exhibit a more significant photo-center displacement, up to a few percent of the stellar radius in the case of RSGs \citep{chiavassa_probing_2022}. 

Since the photo-center is linked to surface convection, it is worth checking for periods in its 
dynamics over the 5 years of observations of Betelgeuse before the dimming. While these periods may overlap with those presented in the previous section, they are likely to capture additional aspects of the phenomena at work.
However, before proceeding to search for periods in the displacement of the photo-center, it is important to address a key issue regarding the interpretation of such images. 
Linear polarization suffers from a $180 ^\circ$ degree ambiguity, as mentioned in \cite{auriere_discovery_2016}. Consequently, our images 
can be 
rotated by $180^\circ$ degree, and the brightness distribution will still fit the observed LSD polarization profiles. 
If each observation were treated independently, the algorithm's solution could be any of the possible ambiguous solutions. To ensure continuity 
between the image series, 
for a given day, we use the brightness distribution of the previous day as the initial point 
of the  fitting iteration. The first image in the series begins its minimization iteration with a random brightness distribution. 
Although we can produce a series of consistent images, it is important to keep in mind that the photo-center displacement computed from the series 
will be affected by the initial image. 
To overcome this issue, we computed
the photo-center displacement from 100 different time series, each starting from a different initial image. This approach aims to recover ensemble properties independent of the choice of the first image. 

\begin{figure}[!h]
    \centering
    \includegraphics[width=0.5\textwidth]{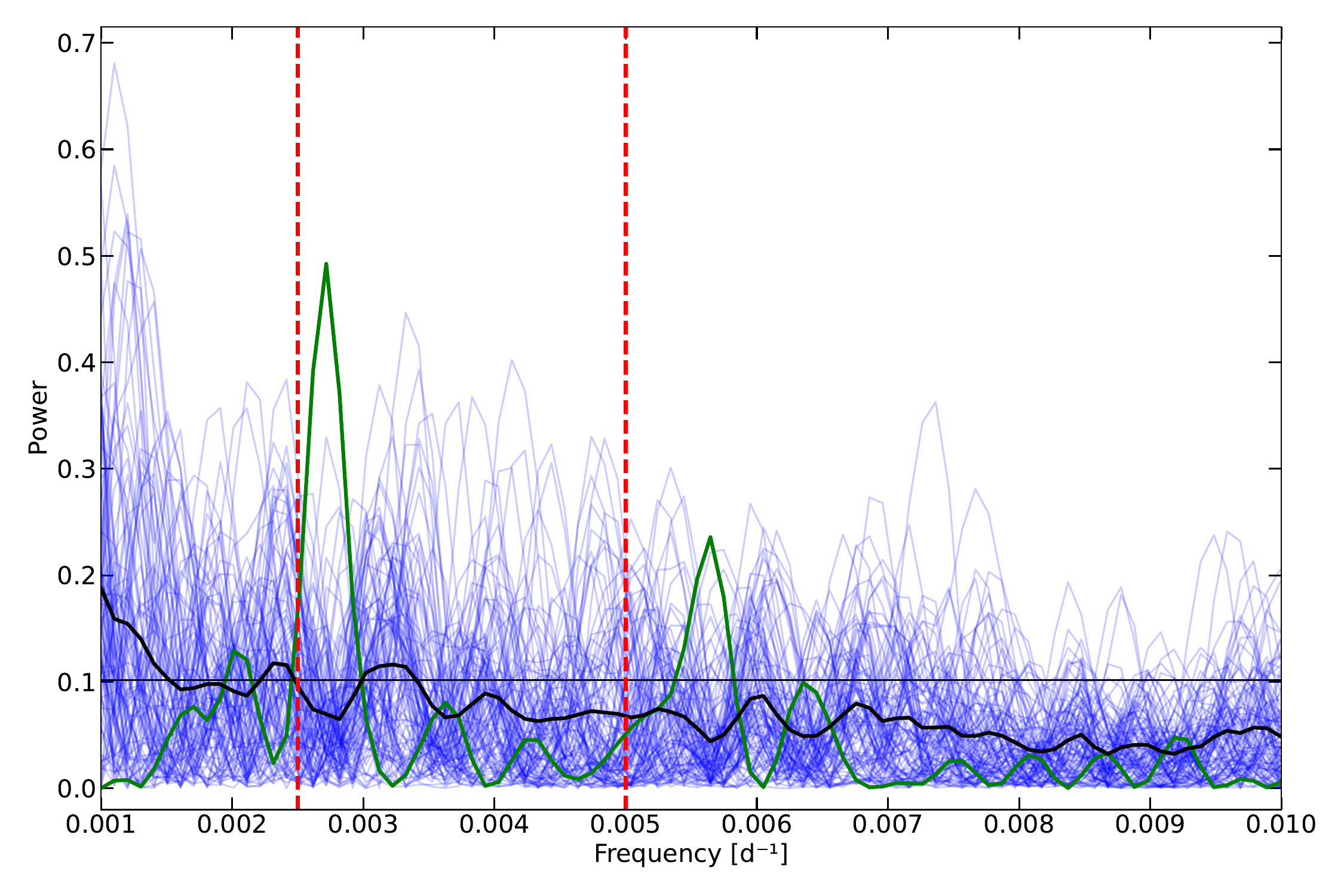}
    \caption{Lomb-Scargle periodogram of the 100 photo-center displacement series of Betelgeuse. Each blue lines correspond to a Lomb-Scargle periodogram of one photo-center
     displacement. The black line is the average of the 100 periodogram. The red dashed lines represent the 400 and 200 d period respectively.
     The green line is the window function. The black horizontal line marks the 1$\sigma$ confidence level.}
    \label{LS photocenter}
\end{figure}

Figure \ref{LS photocenter} shows the Lomb-Scargle periodogram of the photo-center displacement for each of the 100 series (blue lines) and the average Lomb-Scargle periodogram (black line). The red dashed lines indicate the 400 and 
200 d periods, respectively. The black horizontal line represents the 1$\sigma$ confidence level, which was computed from the standard deviation of the average Lomb-Scarlge periodogram. Similar to previous figures, the window function is represented by the green line. Interestingly, we find two peaks around the 400 d period in the periodogram, one at 0.0024 $\mathrm{d^{-1}}$ and one at 0.003 $\mathrm{d^{-1}}$ which are above the 1$\sigma$ confidence level. Although they are not individually significant to confirm the presence of such period, these peaks are consistent with the findings of the previous section. Since polarimetry imaging involves only surface convection, this provides further support for a variability being explained by surface convection alone.

\subsection{Variability of the light curve}

After examining the periods identified by the Lomb-Scargle technique in the polarization data obtained by the TBL over the last years before the dimming, it is worth contextualizing them alongside the periods traditionally identified in light curves over the same period of time. Two aspects 
are of our interest in this comparison: the behavior of the light curve before and after the great dimming.

\begin{figure}[!h]
    \centering
    \includegraphics[width=0.5\textwidth]{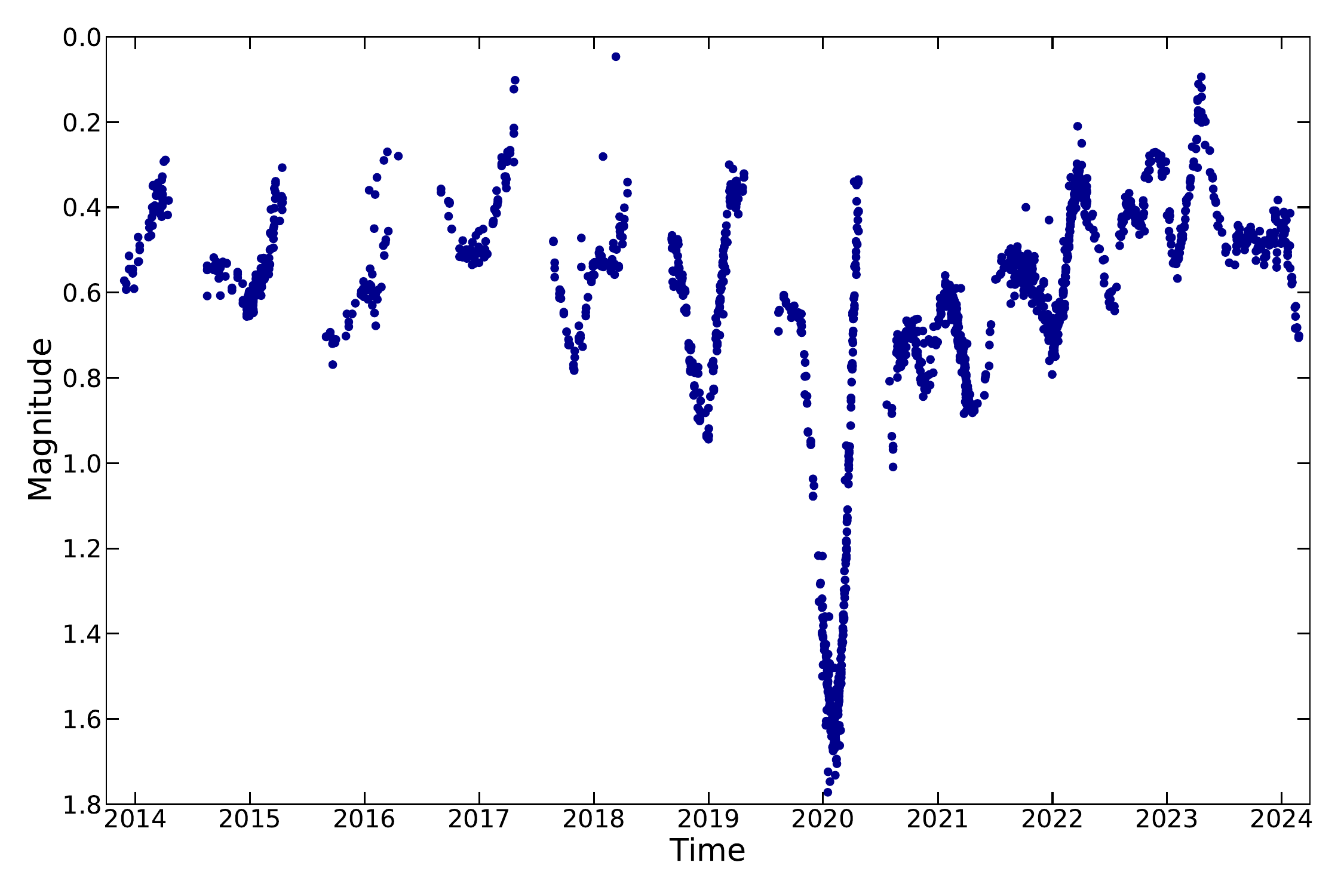}
    \caption{Light curve of Betelgeuse from AAVSO in the V-band.}
    \label{light curve Betelgeuse}
\end{figure}

We have retrieved the light curve of Betelgeuse in the visible from the AAVSO database for the past 10 years. It is shown in Fig. \ref{light curve Betelgeuse}.
Before the great dimming, Betelgeuse's magnitude exhibited variations on a yearly timescale, whereas after the dimming, its variability has shortened. 
Figure \ref{light curve Betelgeuse} clearly illustrates that the 
variability of Betelgeuse now occurs on a timescale shorter than one year. This qualitative change has been pointed out before,
before the great dimming, the primary period of Betelgeuse was approximately 400 d \citep{kiss_variability_2006}, often associated with the fundamental pressure mode. However, after the dimming, this period seems to have vanished, and only timescales shorter than 230 d are visible since \citep{dupree_great_2022}. 
No explanations have been put forth regarding this change of variability after, or perhaps because of the dimming. \

\begin{figure}[!h]
    \centering
    \includegraphics[width=0.5\textwidth]{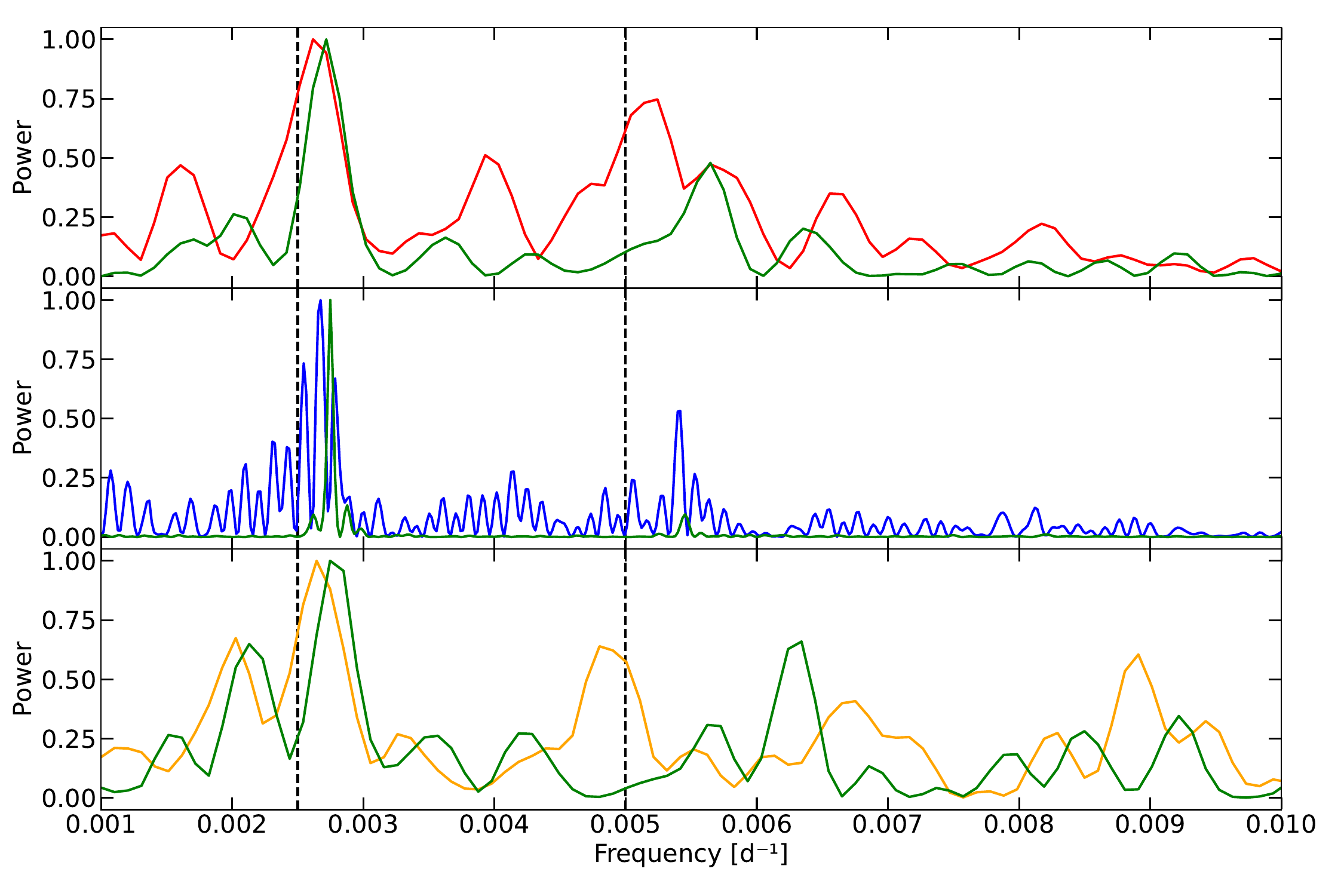}
    \caption{Upper panel: Lomb-Scargle periodogram of the mean intensity profile presented in Fig.\ref{LS intensity} (red line).
    Middle panel: Periodogram of the light curve since 1990 as retrived from the AAVSO database (blue line).
    Lower panel: the orange line represents the periodogram of the light curve from the AAVSO when the dates correspond to the TBL observations. In every panels, the green line marks the associated window function. The 400 d and the 200 d periods are marked by the black dashed lines.}
    \label{Lomb Scargle AAVSO}
\end{figure}

In Fig.\ref{Lomb Scargle AAVSO}, the upper panel is the periodogram of the mean intensity profile (red line) shown in Fig.\ref{LS intensity}. In the middle panel, we computed the periodogram from the light curve of Betelgeuse from 1990 to 2024, represented with the blue line. The orange line in lower panel depicts the periodogram computed exclusively with AAVSO observations made less than 2 days from our observation dates of the TBL, representing a total of 28 observations. In every panels, the green line represents the window function. For the periodogram since 1990, we binned the observations with an interval of 10 days, as Betelgeuse has been more observed in the 21st century \citep{kiss_variability_2006}. This periodogram  recovers the 400 d period and also a period close to 200 d. Other visible peaks are rather 
due to the windowing effect. When focusing on the periodogram derived from the light curve corresponding to the TBL observation dates (lower panel), we recover the 400 d period, which is once again, very close to a peak of the window function. A smaller peak is also present around 200 d. The comparison between this periodogram and the one obtained from Stokes $I$ is important to validate our results. Stokes $I$ is linked to the total amount of light, that is the light curve. Hence, by taking observations of the AAVSO that correspond to the TBL observation dates of Betelgeuse, we should recover the same periods that the ones found in the Stokes $I$ periodogram. Since we found the 400 d and the 200 d periods in both periodograms, this validates our results. In both cases, the peak at 400 d is too close to a peak of the window function to be interpreted as a real peak. On the other hand, the second peak around 200 d, could be interpreted as a real peak. In the future, extended spectropolarimetric observations of Betelgeuse will be necessary to smooth the window function of the TBL and perhaps, disentangle the 400 d period from the window function.

\section{Discussion}
\label{section 4}
 
The presence of the same 
periods in the periodogram derived from the AAVSO and the ones obtained from linear polarization and, considering the time span and sparsity of the TBL data series point to a common origin, that seems rather based upon convective dynamics. Since we associate Stokes $Q$ and $U$ to proxies of convection, we associate the 200 d period to the typical timescales on which the Stokes parameters evolve. 
The LSD profiles of Stokes $Q$ or $U$ undergo slight changes when observing Betelgeuse monthly, but are completely overhauled 
on time scales of several months. The Lomb-Scargle periodograms seem to capture this spectral dynamic at periods of approximately 200 d. Such changes in the profiles are directly linked to the dynamics of the surface,  to  the evolution, and movements 
of convective cells across the visible hemisphere. Therefore, the 200 d period appears to be related to the evolution of the convective patterns 
on the surface of Betelgeuse.  The stronger presence of this peak in Stokes $Q$ compared to $U$ was already pointed out by \cite{mathias_evolution_2018}, and could be interpreted as a temporal situation of how convective patterns have been distributed in the
last years.\

The other period found in the total linear polarization and Stokes $U$, the 330 d periods, is not so far from the 400 days period often mentioned in the literature (e.g \cite{kiss_variability_2006} found a period of $388 \pm 30$ d). It has been shown by \cite{lopez_ariste_convective_2018} that large convective cells can persist for one to two years. Thus, the 330 d period could be associated to the timescale on which the largest granules evolve. This characteristic timescale has also been observed in the numerical simulations and might play an important role in determining stellar distances through the displacement of the photo-center \citep{chiavassa_probing_2022}. Once again, we observe the 330 d periodicity to be more prominent in Stokes $U$ than in $Q$, which may unravel a random situation due to convective motions in the last years. Convective cells could arise in peculiar positions, where the Stokes $U$ profile remains unchanged for months but where Stokes $Q$ could change on shorter timescales. This interpretation aligns with the work of \cite{gray_mass_2008}, who attributed the 400 d period to large convective cells. Whether this 330 d period and the commonly cited 400 d period are one and the same will have to wait for extended spectropolarimetric observations with the TBL that smooth the windowing function around these frequencies.  \

Regarding the Stokes $I$ profile, a peak closer to the 400 d period appears, but is once again uncomfortably close to a peak of the window function of the TBL, making it difficult to trust. A secondary peak close to 200 d is also present in the Stokes $I$ profile, that we fully assimilate to the one found in the Stokes parameters, and that we attribute to convective timescales.\\

Before summing up the results of our work, we can afford to propose an explanation for the variability of Betelgeuse after the great dimming with surface convection. RSGs experience mass loss event, where plasma is ejected into the interstellar medium \citep{josselin_atmospheric_2007}. These events have been inferred in RSG such as $\mu$~Cep at photospheric heights, where \cite{lopez_ariste_height_2023} found an excess of linear polarization beyond the limb of the star, attributing it to rising plumes of plasma in the back hemisphere of the star. In the case of Betelgeuse, the excess of linear polarization beyond the limb is rare and not as pronounced as those observed in $\mu$~Cep. They do not appear to contribute to the variability of the star. However, since the great dimming, the variability of Betelgeuse has changed. This suggests that the great dimming differed significantly from the usual mass loss event, that it hustled the dynamics of the photosphere. \

In a scenario where the variability of Betelgeuse is primarily governed by convection, the visible periods of such variability were associated, before the great dimming, to the typical timescales at which convection occurs: approximately 200 d for the small structures and roughly 350 d for the large structures. The great dimming affected the behavior of the photosphere in
such a way that the largest structures no longer evolved on a
timescale around 350 d, but rather on the shorter timescales of
the smallest convective structures, around 200 d. We hypothesize that after the great dimming event, the photosphere has not
yet returned to equilibrium, and the largest convective structures
have been disrupted by the turbulent motions present of the photosphere in this period. This would explain why identifying the
longer periods after the great dimming has been hard, as \cite{dupree_great_2022} noticed. We may expect that the 400 d period will
gradually reappear in the coming years as the photosphere returns to equilibrium. However speculative, this scenario provides
a rough explanation of how convective activity could lead to a
change in variability since the great dimming that future observations and research will confirm or infirm.

\section{Conclusion}
\label{Conclusion}

The main result presented in this work is that the same periods identified in the light curve of Betelgeuse are also observed in the intensity and polarization
spectra measured with the TBL. Traditionally, periods in the light curve have been associated to pulsations due to pressure modes. However, the interpretation of the linear polarization profiles is made in terms of convective structures in the atmosphere of Betelgeuse. We propose that the true reason for the observed periodicities in both the light curve and spectropolarimetric observations lies within these convective structures and their temporal evolution. This is the main conclusion of our work: since we are able to recover the different periods in the linear polarization profiles, we can exclude pulsation phenomena as the explanation for those periodicities in the
light curve. This conclusion concurs with the scenario proposed by \cite{gray_mass_2008}, attributing the variability of Betelgeuse to surface convection. We further find hints in the polarized spectra that lead us to associate the 200 d periodicity to convective dynamics, while the 400 d periodicity may be linked to convective timescales of the largest granules.

Although of secondary importance, we can speculate on the phenomena underlying the change in variability before and after the great dimming event. We hypothesize that the great dimming event disrupted the dynamics of the photosphere, leading to the destruction of the largest granules by the turbulent motions present in the non-equilibrium of the photosphere after such event. Since this happened, the variability of Betelgeuse has quickened in adequacy with the timescale of smaller convective cells, around 200 d. If this scenario is correct, we shall expect in the coming future, the re-emergence of the 400 d variability as the dominant period once the photosphere returns to some form of equilibrium. Longer time 
series of spectropolarimetric observations of Betelgeuse will be necessary to provide further insights into this phenomenon.

\begin{acknowledgements}
    This work was supported by the "Programme National de Physique Stellaire" (PNPS) of CNRS/INSU co-funded by CEA and CNES.
    We acknowledge support from the French National Research Agency (ANR)
    funded project PEPPER (ANR-20-CE31-0002).
    We acknowledge the observers of the AAVSO International Database who provided more data than enough to produce this work.
    
    \end{acknowledgements}
    
    \bibliographystyle{aa}

    \bibliography{main}
\end{document}